# Influences of the Transverse Motions of the Particles to the Recombination Rate of a Co-propagating Electron-ion System

G. Wang

Abstract

For a system with the ion beam co-propagating with the electron beam, such as a traditional electron cooler or a Coherent electron Cooler (CeC), the recombination rate is an important observable for matching the energy of the electrons with the ions. In this work, we have developed the analytical expressions to investigate how the recombination rate depends on the energy difference of the two beams, with the influences from the transverse motions of the particles being considered. The analytical results are then used to analyze the measured recombination data collected during the CeC experiment in run 21.

## I. Recombination rate for Gaussian transverse velocity distribution

According to eq. (2) of [1], the recombination rate is given by

$$\alpha_r = \frac{\int_{-\infty}^{\infty} d^3v_i d^3v_e f_e(v_e) f_I(v_i) |\vec{v}_e - \vec{v}_i| \sigma(|\vec{v}_e - \vec{v}_i|)}{\int_{-\infty}^{\infty} d^3v_i d^3v_e f_e(v_e) f_I(v_i)}, \quad (1)$$

where $\sigma(|\vec{v}_e - \vec{v}_i|)$ is the recombination cross section which depends on the relative velocity of an ion with respect to an electron, and $f_e(v_e)$ and $f_I(v_i)$ are velocity distributions of the electrons and ions. We assume that the velocity distribution of electrons is

$$f_e(v_e) = \frac{1}{2\pi\beta_{e,\perp}^2} \exp\left(-\frac{v_{e,x}^2 + v_{e,y}^2}{2\beta_{e,\perp}^2}\right) f_{e,z}(v_{e,z}), \quad (2)$$

and that of ions is

$$f_I(v_i) = \frac{1}{2\pi\beta_{i,\perp}^2} \exp\left(-\frac{v_{i,x}^2 + v_{i,y}^2}{2\beta_{i,\perp}^2}\right) f_{i,z}(v_{i,z}), \quad (3)$$

with $\beta_{e,\perp}$ and $\beta_{i,\perp}$ being the transverse velocity spread of the electrons and the ions. Using the following relation

$$\exp\left(-\frac{v_{i,x}^2}{2\beta_{i,\perp}^2} - \frac{v_{e,x}^2}{2\beta_{e,\perp}^2}\right) = \exp\left\{-\mu_\perp \left(v_{i,x} + \frac{v_x}{2\mu\beta_{e,\perp}^2}\right)^2\right\} \exp\left[\frac{-v_x^2}{2(\beta_{e,\perp}^2 + \beta_{i,\perp}^2)}\right], \quad (4)$$

where $\mu_\perp \equiv \dfrac{1}{2\beta_{i,\perp}^2} + \dfrac{1}{2\beta_{e,\perp}^2}$ and $\vec{v} \equiv \vec{v}_e - \vec{v}_i$, eq. (1) can be written as

$$\alpha_r = \dfrac{\int_{-\infty}^{\infty} d^3v\, v\sigma(v) \exp\left[-m_e \dfrac{v_x^2 + v_y^2}{2kT_{ei}}\right] \int_{-\infty}^{\infty} f_{e,z}(v_z + v_{i,z}) f_{i,z}(v_{i,z}) dv_{i,z}}{2\pi \dfrac{kT_{ei}}{m_e} \int_{-\infty}^{\infty} f_{e,z}(v_z + v_{i,z}) f_{i,z}(v_{i,z}) dv_z dv_{i,z}}, \tag{5}$$

where we defined the effective temperature parameter, $T_{ei}$, as

$$kT_{ei} = m_e \left(\beta_{e,\perp}^2 + \beta_{i,\perp}^2\right) = \dfrac{1}{2} m_e \left(\beta_{e,x}^2 + \beta_{e,y}^2 + \beta_{i,x}^2 + \beta_{i,y}^2\right). \tag{6}$$

## II. Longitudinally cold electrons and ions

Typically, the longitudinal velocity spread in the beam frame is much smaller than the transverse velocity spread and hence we can take the delta function for the longitudinal velocity distribution, i.e.

$$f_{e,z}(v_{e,z}) = \delta(v_{e,z} - v_{z0}), \tag{7}$$

and

$$f_{i,z}(v_{i,z}) = \delta(v_{i,z}). \tag{8}$$

Inserting eq. (7) and (8) into eq. (5) yields

$$\alpha_r = \dfrac{1}{2\pi kT_{ei}/m_e} \int_{-\infty}^{\infty} d^3v\, v\sigma(v) \exp\left[-\dfrac{v_x^2 + v_y^2}{2kT_{ei}/m_e}\right] \delta(v_z - v_{z0}). \tag{9}$$

According to eq. (23) of [2], the recombination cross section for an electron moving with velocity $(v_x, v_y, v_z)$ with respect to the ion is

$$\sigma(v_x, v_y, v_z) = A \dfrac{2h\nu_0}{m_e(v_x^2 + v_y^2 + v_z^2)} \left[\ln\left(\sqrt{\dfrac{2h\nu_0}{m_e(v_x^2 + v_y^2 + v_z^2)}}\right) + \gamma_1 + \gamma_2 \left(\dfrac{m_e(v_x^2 + v_y^2 + v_z^2)}{2h\nu_0}\right)^{1/3}\right] \tag{10}$$

where $A = 2.11 \times 10^{-22}\, cm^2$, $h\nu_0 = Z^2\alpha^2 m_e/2 = Z^2 \times 13.6\, eV$, $Z = 79$, $\gamma_1 = 0.1402$ and $\gamma_2 = 0.525$. Inserting eq. (10) into eq. (9) yields

$$\alpha_r = \frac{m_e}{2\pi kT_{ei}} \int_{-\infty}^{\infty} d^3v v \sigma(v) \exp\left[-\frac{v_x^2+v_y^2}{2kT_{ei}/m_e}\right] \delta(v_z - v_{z0})$$

$$= Ac\frac{h\nu_0}{kT_{ei}}\sqrt{\frac{2h\nu_0}{m_e c^2}} \exp\left(\frac{v_{z0}^2}{2kT_{ei}/m_e}\right) \int_{\frac{m_e v_{z0}^2}{2h\nu_0}}^{\infty} \frac{1}{\sqrt{y}} \left[\ln\left(\sqrt{\frac{1}{y}}\right) + \gamma_1 + \gamma_2 y^{1/3}\right] \exp\left(-\frac{h\nu_0}{kT_{ei}} y\right) dy, \quad (11)$$

or

$$\bar{\alpha}_r \equiv \frac{\alpha_r}{Ac} = \frac{h\nu_0}{kT_{ei}}\sqrt{\frac{2h\nu_0}{m_e c^2}} \exp\left(\frac{m_e c^2}{2kT_{ei}}\delta_{z0}^2\right) \int_{\frac{m_e c^2 \delta_{z0}^2}{2h\nu_0}}^{\infty} \frac{1}{\sqrt{y}} \left[-\frac{1}{2}\ln y + \gamma_1 + \gamma_2 y^{1/3}\right] \exp\left(-\frac{h\nu_0}{kT_{ei}} y\right) dy. \quad (12)$$

For CeC experiment, $\beta_{e,x} \approx 2.6\times 10^6\,m/s \sim 8\times 10^6\,m/s$, $\beta_{e,z} \approx 1.5\times 10^5\,m/s$, $\beta_{i,x} \approx 1.2\times 10^6\,m/s$ and $\beta_{i,z} \approx 3.9\times 10^5\,m/s$. Fig. 1 compares the recombination curves as calculated from eq. (12) and that measured from CeC experiment for 0.9 mrad of electron beam angular spread, i.e. $\beta_{e,x} \approx 8\times 10^6\,m/s$, showing the width of the recombination curve from calculation is comparable to that from the measurement. The abscissa of fig. 1 (left) and all the following figures are

$$v_{z0}/c = (\langle v_{e,z}\rangle - \langle v_{i,z}\rangle)/c, \quad (13)$$

where $\langle ...\rangle$ represents averaging over the beam distribution, $v_{e,z}$ is the longitudinal velocity of the electrons in the beam frame and $v_{i,z}$ is the longitudinal velocity of the ions in the beam frame. In the ultra-relativistic limit, for an electron moving with velocity $\vec{v}_e$ and scattering off an ion moving with velocity $\vec{v}_i$ in the co-moving frame, one can derive the difference in their relativistic factor $\gamma_e$ and $\gamma_i$ in the lab frame is given by

$$\frac{\gamma_e - \gamma_i}{\gamma} \approx \frac{v_{e,z} - v_{i,z}}{c} + \frac{1}{2}\gamma^2\left[\theta_{e,x}^2 + \theta_{e,y}^2 - \theta_{i,x}^2 - \theta_{i,y}^2\right] \quad (14)$$

(see Appendix), where $\gamma_e$ and $\gamma_i$ are the relativistic energy factor of the two particles, and $\theta$ is the transverse angle in the lab frame. Averaging eq. (14) over the distribution of the two beams yields

$$v_{z0}/c = (\langle v_{e,z}\rangle - \langle v_{i,z}\rangle)/c \approx (\langle\gamma_e\rangle - \langle\gamma_i\rangle)/\langle\gamma_i\rangle - \frac{1}{2}\gamma^2\left(\langle\theta_{e,\perp}^2\rangle - \langle\theta_{i,\perp}^2\rangle\right), \quad (15)$$

with $\langle\theta_{e,\perp}^2\rangle = \langle\theta_{e,x}^2\rangle + \langle\theta_{e,y}^2\rangle$ and $\langle\theta_{i,\perp}^2\rangle = \langle\theta_{i,x}^2\rangle + \langle\theta_{i,y}^2\rangle$. For small angular spread of the two beams, the abscissa of fig. 1 (left) is approximately equal to the relative difference of the relativistic $\gamma$ factor of the two beams.

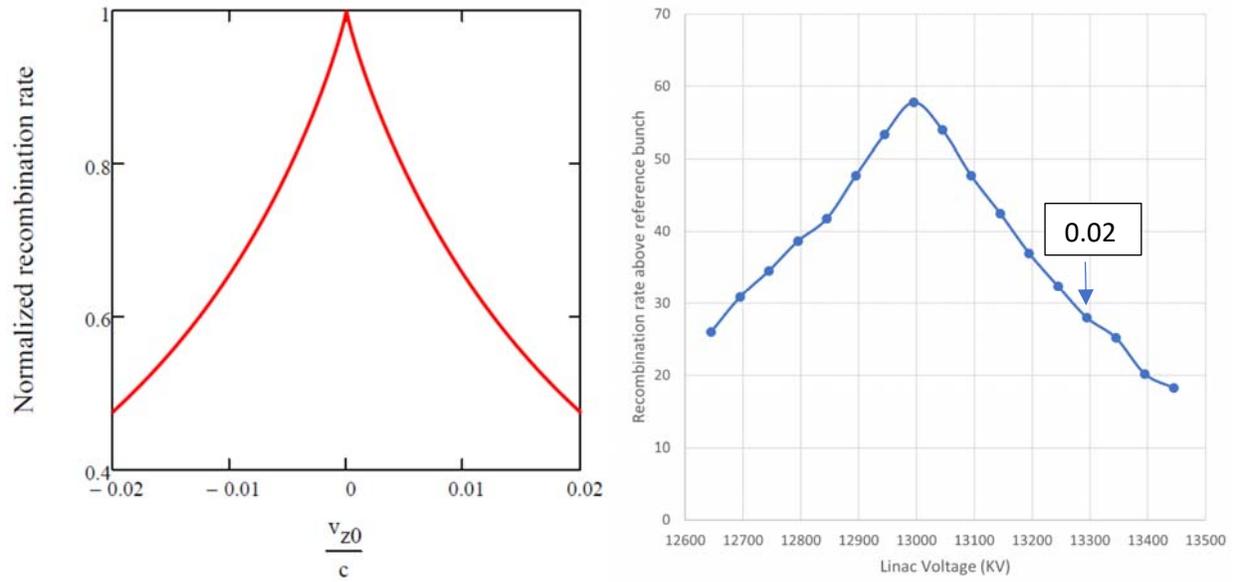

Figure 1: comparison of recombination rate as predicted from eq. (12) and that measured in CeC experiment. (Left) normalized recombination rate as a function of $v_{z0}/c$ for transverse velocity spread of $\beta_{e,x} = \beta_{e,y} = 8\times 10^6 m/s$ (about 0.93 mrad angular spread in lab frame) and $\beta_{i,x} = \beta_{i,y} = 1.2\times 10^6 m/s$; (Right) measured recombination rate in CeC experiment.

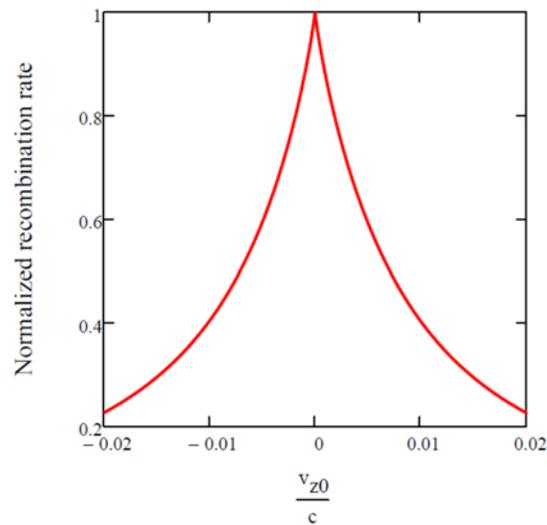

Figure 2: normalized recombination rate as a function of $v_{z0}/c$ of the electrons for transverse velocity spread of $\beta_x = \beta_y = 2.6\times 10^6 m/s$ (about 0.3 mrad angular spread in lab frame) as calculated from eq. (12).

Fig.2 shows that if there is no additional angular spread coming from misalignment, the 0.3 mrad angular spread from designed lattice alone will result in a recombination curve which is a factor of 2 narrower than what measured from the experiment.

## III. Influences of orbital angle due to misalignment

If the electrons merges into ions with an angle, the average transverse velocity of the electrons in the beam frame is nonzero. In this case, eq. (11) becomes

$$\alpha_r = \frac{m_e}{2\pi k T_{ei}} \int_{-\infty}^{\infty} d^3 v \, v \sigma(v) \exp\left[-\frac{(v_x - v_{x0})^2 + v_y^2}{2kT_{ei}/m_e}\right] \delta(v_z - v_{z0})$$

$$= \frac{Ach\nu_0}{kT_{ei}} \sqrt{\frac{2h\nu_0}{m_e c^2}} \exp\left(\frac{v_{z0}^2 - v_{x0}^2}{2kT_{ei}/m_e}\right) \quad , \quad (16)$$

$$\times \int_{\frac{m_e v_{z0}^2}{2h\nu_0}}^{\infty} \frac{1}{\sqrt{y}} \left[\ln\left(\sqrt{\frac{1}{y}}\right) + \gamma_1 + \gamma_2 y^{1/3}\right] I_0\left(\frac{v_{x0}\sqrt{2m_e h\nu_0}}{kT_{ei}}\sqrt{y - \frac{m_e v_{z0}^2}{2h\nu_0}}\right) \exp\left[-\frac{h\nu_0 y}{kT_{ei}}\right] dy$$

where $v_{x0}$ is the average horizontal velocity of the electrons with respect to the ions and $I_0(x)$ is the modified Bessel function.

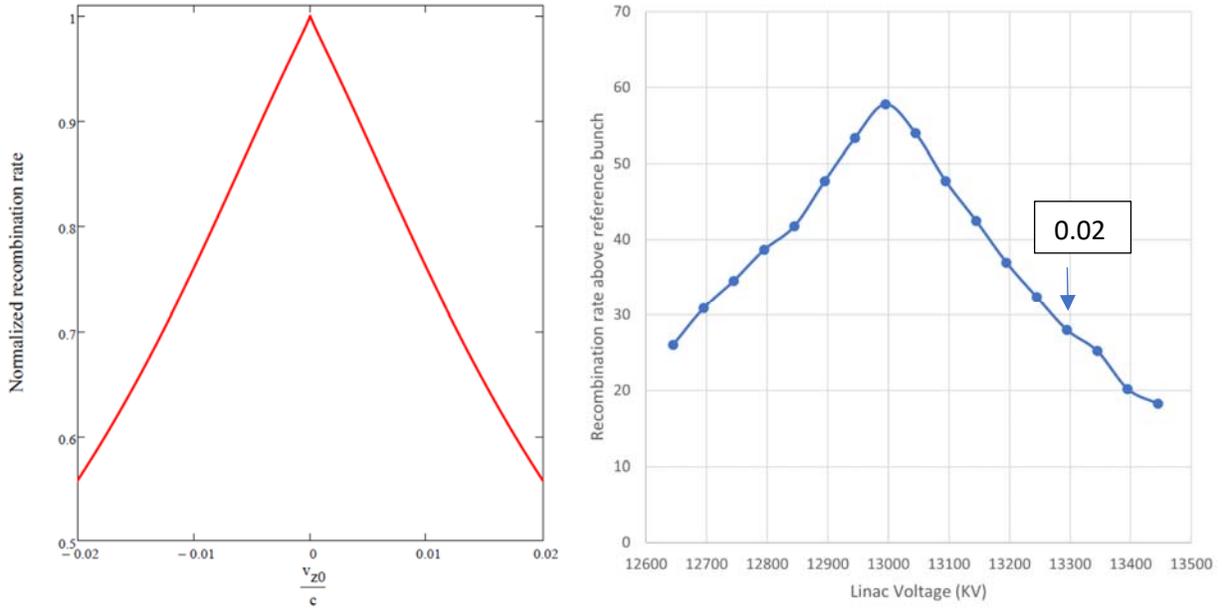

Figure 3: normalized recombination rate as a function of $v_{z0}/c$ for transverse velocity spread of $\beta_x = \beta_y = 2.6\times 10^6 \, m/s$ (about 0.3 mrad angular spread in lab frame) and transverse average velocity of $v_{x0} = 7\times 10^6 \, m/s$ (about 0.8 mrad orbital angle) as calculated from eq. (16).

As shown in fig. 3, an orbital angle of 0.8 mrad with angular spread of 0.3 mrad is enough to explain the measured recombination curve (fig. 3 right) from CeC experiment. It is worth noting that in the presence of an orbital angle, $\theta_x = v_{x0}/(\gamma\beta c)$, the relation between the energy deviation of the two beams and the abscissa of fig. 3 (left) becomes

$$v_{z0}/c = (\langle v_{e,z}\rangle - \langle v_{i,z}\rangle)/c \approx (\langle \gamma_e\rangle - \langle \gamma_i\rangle)/\langle \gamma_i\rangle - \frac{1}{2}\gamma^2\left(\langle \theta_{e,\perp}^2\rangle - \langle \theta_{i,\perp}^2\rangle + \theta_x^2\right),$$

i.e. the orbital angles due to misalignment introduces additional shifts of the electrons' energy for reaching the maximal recombination rate.

### IV. Gaussian distribution of longitudinal velocity

In the previous sections, we assume that, in the co-moving frame of the ions, the longitudinal velocity spreads of the electrons and the ions are negligible compared to their transverse velocity spreads. To investigate how the longitudinal velocity spread may affect the recombination rate, we take the Gaussian longitudinal velocity distribution, i.e.

$$f_{e,z}(v_{e,z}) = \frac{1}{\sqrt{2\pi}\beta_{e,z}}\exp\left(-\frac{(v_{e,z}-v_{z0})^2}{2\beta_{e,z}^2}\right), \tag{17}$$

and

$$f_{i,z}(v_{i,z}) = \frac{1}{\sqrt{2\pi}\beta_{i,z}}\exp\left(-\frac{v_{i,z}^2}{2\beta_{i,z}^2}\right), \tag{18}$$

where $\beta_{e,z}$ and $\beta_{i,z}$ are the longitudinal velocity spread of the electrons and ions in the co-moving frame of the ions. Inserting eq. (10), eq. (17) and eq. (18) into eq. (5) yields

$$\alpha_r = \frac{\int_{-\infty}^{\infty} d^3v\, v\sigma(v)\exp\left[-m_e\frac{v_x^2+v_y^2}{2kT_{ei}}\right]\int_{-\infty}^{\infty}\exp\left(-\frac{(\tilde{v}_z+v_{i,z})^2}{2\beta_{e,z}^2}\right)\exp\left(-\frac{v_{i,z}^2}{2\beta_{i,z}^2}\right)dv_{i,z}}{2\pi\frac{kT_{ei}}{m_e}\int_{-\infty}^{\infty}\int_{-\infty}^{\infty}\exp\left(-\frac{(\tilde{v}_z+v_{i,z})^2}{2\beta_{e,z}^2}\right)\exp\left(-\frac{v_{i,z}^2}{2\beta_{i,z}^2}\right)dv_z dv_{i,z}}$$

$$= \frac{Ach v_0}{kT_{ei}\eta}\sqrt{\frac{2h v_0}{m_e c^2}}\exp\left(\frac{m_e v_{z0}^2}{4kT_z}\left(\frac{1}{\eta^2}-1\right)\right)\int_0^{\infty}(-\ln y + \gamma_1 + \gamma_2 y^{2/3})\exp\left(-\frac{h v_0}{kT_{ei}}y^2\right)$$

$$\times\left\{\text{Erf}\left[\eta\sqrt{\frac{h v_0}{2kT_z}}\left(y+\sqrt{\frac{m_e}{2h v_0}}\frac{v_{z0}}{\eta^2}\right)\right] + \text{Erf}\left[\eta\sqrt{\frac{h v_0}{2kT_z}}\left(y-\frac{v_{z0}}{\eta^2}\sqrt{\frac{m_e}{2h v_0}}\right)\right]\right\}dy \tag{19}$$

where $\tilde{v}_z = v_z - v_{z0}$, $\eta = \sqrt{\frac{T_{ei}-2T_z}{T_{ei}}}$ and $kT_z = \frac{1}{2}m_e(\beta_{e,z}^2 + \beta_{i,z}^2)$. It is worth noting that for $T_{ei} < 2T_z$, $\eta$ is imaginary but according to the relation, $erf(i\cdot x) = i\cdot erfi(x)$, the error function in eq. (19)

also returns an imaginary value and hence the expression is real and still valid for calculating the recombination rate. Eq. (19) has a singularity at $T_{ei} = 2T_z$ and in this case, the following expression can be used

$$\alpha_r = Ac \frac{h\nu_0}{m_e c v_{z0}} \sqrt{\frac{2h\nu_0}{\pi k T_z}} \exp\left(\frac{-m_e v_{z0}^2}{4kT_z}\right)$$

$$\times \int_0^\infty \left[-\frac{1}{2}\ln y + \gamma_1 + \gamma_2 y^{1/3}\right] \frac{1}{\sqrt{y}} \exp\left(-\frac{h\nu_0 y}{2kT_z}\right) \sinh\left(\frac{v_{z0}\sqrt{m_e h\nu_0 y}}{\sqrt{2kT_z}}\right) dy \quad (20)$$

Fig. 4 shows the recombination curve as calculated from eq. (19) and compares it with results from eq. (12) where energy spread is neglected. The red curves in fig. 4 shows the results for the nominal energy spreads and the blue curves shows the results with the energy spread of both beams reduced by a factor of 100. The green dash curves in fig. 4 are calculated with eq. (12), showing good agreement between eq. (12) and eq. (19) at the small energy spread limit.

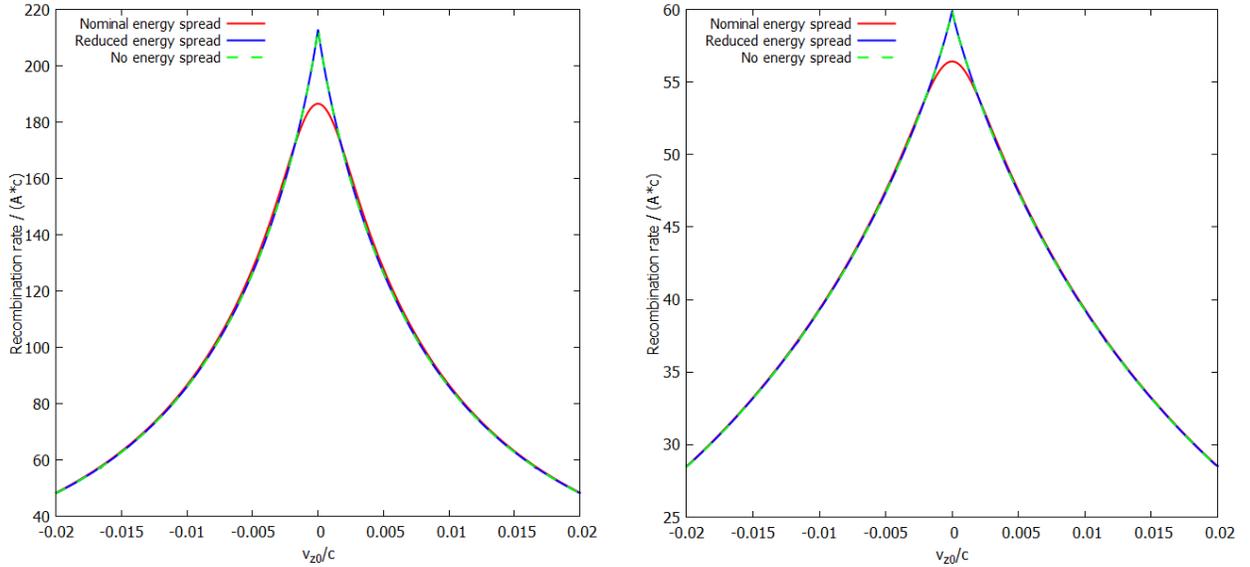

Figure 4: (Left) recombination rate as a function of $v_{z0}/c$ for transverse velocity spread of $\beta_{e,x} = \beta_{e,y} = 2.6\times10^6 m/s$ (about 0.3 mrad angular spread in lab frame) and $\beta_{i,x} = 1.2\times10^6 m/s$. (Right) same as (Left) but with $\beta_{e,x} = \beta_{e,y} = 8\times10^6 m/s$ (about 0.93 mrad angular spread in lab frame). For the red curves, the longitudinal velocity spread of $\beta_{e,z} = 1.5\times10^5 m/s$ (5e-4 rms energy spread) and $\beta_{i,z} = 3.9\times10^5 m/s$ have been assumed to generate the plots from eq. (19). The blue curve is also generated from eq. (19) but with 100-fold smaller energy spread for both beams. The green dash curve is generated from the previous results for zero energy spread, i.e. eq. (12).

Fig. 5 shows the recombination rate with angular spread of the two beams reduced by a factor of 10 (blue) and a factor of 20 (black). For small angular spread, the recombination curve is dominated by the energy spread of the two beams and resembles a Gaussian shape when the energy difference between the electrons and the ions is small. For larger $v_{0z}$, the recombination rate is dominated by the energy difference between the two beams and as shown in fig. 5, all three curves give similar recombination rate at $v_{0z} \sim 0.02c$.

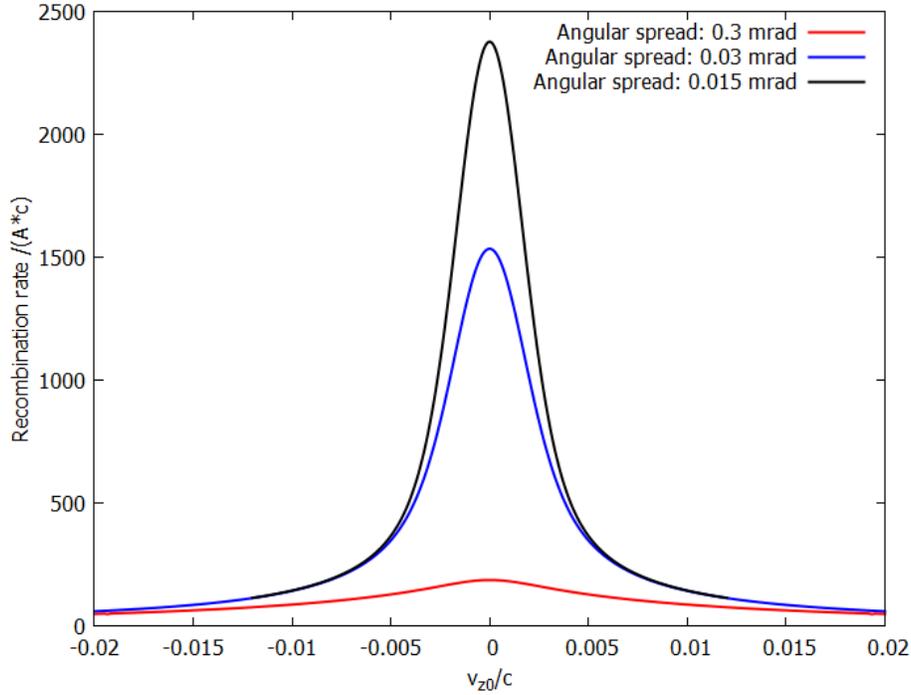

Figure 5: recombination rate as a function of $v_{0z}/c$ for various transverse velocity spread of the two beams. The red curve is for the designed transverse velocity spread, i.e. $\beta_{e,x} = \beta_{e,y} = 2.6\times10^6 m/s$ (0.3 mrad angular spread in the lab frame) and $\beta_{i,x} = \beta_{i,y} = 1.2\times10^6 m/s$. The blue curve is for 10-fold less angular spread for both beams and the black curve is for 20- fold less angular spread for both beams. The longitudinal velocity spread of $\beta_{e,z} = 1.5\times10^5 m/s$ (5e-4 rms energy spread) and $\beta_{i,z} = 3.9\times10^5 m/s$ (1.3e-3 rms energy spread) are used in generating the plots with eq. (19). The red curve is amplified by a factor of 5 for visibility.

## V. Summary

In this work, analytical expressions have been derived to calculate the recombination rate for an ion beam co-moving with an electron beam with their energy close to each other. A general expression, eq. (19), is derived to calculate the recombination rate with arbitrary angular spread and energy spread of the two beams. For negligible energy spreads, eq. (19) reduces to a simpler expression, eq. (12). The influence of an orbital angle due to the misalignment of the electron beam is also considered for the case of zero energy spreads and an analytical expression, eq. (16), is derived to calculated the recombination rate.

## ACKNOLEDGMENTS

The work is inspired by a talk given by P. Thieberger and has greatly benefitted from the discussion with P. Thieberger, V. N. Litvinenko and D. Kayran. The author is grateful for D. Kayran to point out a mistake in the previous version of the manuscript. This work was supported by Brookhaven Science Associates, LLC under Contract No. DE-SC0012704 with the U.S. Department of Energy.

**APPENDIX A:**

In the co-moving frame (the frame where average velocity of the electrons and the ions equal to zero), the 4D momentum of the two particles are

$$p_e = \left(p_x, 0, p_z, \sqrt{m_e^2 c^2 + p_x^2 + p_z^2}\right) \quad (A1)$$

$$p_i = \left(p_{i,x}, 0, p_{i,z}, \sqrt{m_i^2 c^2 + p_{i,x}^2 + p_{i,z}^2}\right) \quad (A2)$$

In the lab frame, after boost in longitudinal direction, the 4D momentum become

$$p_{i,lab} = \left(p_{i,x}, 0, \gamma p_{i,z} + \gamma\beta\sqrt{m_i^2 c^2 + p_{i,x}^2 + p_{i,z}^2}, \gamma\sqrt{m_i^2 c^2 + p_{i,x}^2 + p_{i,z}^2} + \beta\gamma p_{i,z}\right) \quad (A3)$$

$$p_{e,lab} = \left(p_x, 0, \gamma_i p_z + \gamma_i \beta_i \sqrt{m_e^2 c^2 + p_x^2 + p_z^2}, \gamma_i \sqrt{m_e^2 c^2 + p_x^2 + p_z^2} + \beta_i \gamma_i p_z\right) \quad (A4)$$

The energy of the electron in the lab frame is

$$E_e = \gamma c\left[\sqrt{m_e^2 c^2 + p_x^2 + p_z^2} + \beta p_z c\right], \quad (A5)$$

and hence

$$\gamma_e = \frac{E_e}{m_e c^2} = \gamma\left[\sqrt{1 + \frac{v_{e,x}^2}{c^2} + \frac{v_{e,z}^2}{c^2}} + \beta \frac{v_{e,z}}{c}\right] \approx \gamma\left[1 + \frac{v_{e,x}^2}{2c^2} + \frac{v_{e,z}^2}{2c^2} + \beta \frac{v_{e,z}}{c}\right] \quad (A6)$$

The energy of the ion in the lab frame is

$$E_i = \gamma c\left[\sqrt{m_i^2 c^2 + p_{i,x}^2 + p_{i,z}^2} + \beta p_{i,z} c\right], \quad (A7)$$

and hence

$$\gamma_i = \frac{E_i}{m_i c^2} = \gamma\left[\sqrt{1 + \frac{v_{i,x}^2}{c^2} + \frac{v_{i,z}^2}{c^2}} + \beta \frac{v_{i,z}}{c}\right] = \gamma\left[1 + \frac{v_{i,x}^2}{2c^2} + \frac{v_{i,z}^2}{2c^2} + \beta \frac{v_{i,z}}{c}\right]. \quad (A8)$$

Hence, the relative difference in the $\gamma$ factors is

$$\begin{aligned}
\frac{\gamma_e - \gamma_i}{\gamma} &\approx \frac{v_{e,z} - v_{i,z}}{c} + \frac{1}{2}\left[\frac{v_{e,x}^2}{c^2} + \frac{v_{e,y}^2}{c^2} - \frac{v_{i,x}^2}{c^2} - \frac{v_{i,y}^2}{c^2}\right] \\
&= \frac{v_{e,z} - v_{i,z}}{c} + \frac{1}{2}\gamma^2\left[\frac{p_{e,x}^2}{m^2\gamma^2 c^2} + \frac{p_{e,y}^2}{m^2\gamma^2 c^2} - \frac{p_{i,x}^2}{m^2\gamma^2 c^2} - \frac{p_{i,y}^2}{m^2\gamma^2 c^2}\right]. \\
&\approx \frac{v_{e,z} - v_{i,z}}{c} + \frac{1}{2}\gamma^2\left[\theta_{e,x}^2 + \theta_{e,y}^2 - \theta_{i,x}^2 - \theta_{i,y}^2\right]
\end{aligned} \qquad (A9)$$